# Photonic Crystal Optical Tweezers

Benjamin K. Wilson[1], Tim Mentele[2], Stephanie Bachar[3], Emily Knouf[4], Ausra Bendoraite[4], Muneesh Tewari[4,5], Suzie H. Pun[2], and Lih Y. Lin[1,*]

[1]Department of Electric Engineering, University of Washington, Seattle, WA 98195

[2]Department of Bioengineering, University of Washington, Seattle, WA 98195

[3]Department of Biological Engineering, Massachusetts Institute of Technology, Boston, MA 02139

[4]Human Biology Division, Fred Hutchinson Cancer Research Center, Seattle, WA 98109

[5]Clinical Research and Public Health Sciences Divisions, Fred Hutchinson Cancer Research Center, Seattle, WA 98109

*e-mail: lylin@uw.edu

**Non-invasive optical manipulation of particles has emerged as a powerful and versatile tool for biological study and nanotechnology. In particular, trapping and rotation of cells, cell nuclei and sub-micron particles enables unique functionality for various applications such as tissue engineering[1,2], cancer research[3] and nanofabrication[4]. We propose and demonstrate a purely optical approach to rotate and align particles using the interaction of polarized light with photonic crystal nanostructures to generate enhanced trapping force. With a weakly focused laser beam we observed efficient trapping and transportation of polystyrene beads with sizes ranging from 10 μm down to 190 nm as well as cancer cell nuclei. In addition, we demonstrated alignment of non-spherical particles using a 1-D photonic crystal structure. Bacterial cells were trapped, rotated and aligned with optical intensity as low as 17 μW/μm$^2$. Finite-difference time domain (FDTD) simulations of the optical near-field and far-field above the photonic crystal nanostructure reveal the origins for the observed results. This approach can be extended to using 2-D photonic crystal nanostructures for full rotation control.**

Since the inception of optical tweezers[5], researchers have been exploring new ways to increase their performance and functionality[6] while reducing the required optical intensity to achieve sufficient trapping force. Decreased optical intensity is essential for biological applications where intense optical radiation may be a concern[7-10]. Direct conversion from optical energy to mechanical energy has low efficiency, which is the underlying reason for high optical intensity requirement of optical tweezers. The challenge becomes more severe as the particle size decreases to sub-micron regime. The requirement on optical intensity can be reduced by combining conventional dielectrophoresis with optically induced conductivity change[11]. On the other hand, being able to achieve this goal through pure optical means can preserve the full configurability of optical tweezers with less system complexity, and impose no constraint on the conductivity of the sample solution. The primary approach along this direction entails optical near-field enhancement through interaction with plasmonic nanostructures to increase the local intensity and gradient of the field[12]. This has been implemented using patterned metal nanostructures[13-15], dimer nanoantennas [16], metal nanorods[17] under direct Gaussian beam illumination or through total internal reflection configurations. Enhanced optical near field can also be generated by coupling light into sub-wavelength silicon slot waveguides[18]. While such methods have shown impressive results in increasing trapping force, trapping is limited to specific plasmonic "hot spots" or waveguide slot locations, thus limiting their ability to simply holding particles in predetermined positions patterned onto a surface. This diminishes the functionality of conventional optical tweezers, where particles can be translated in any direction to arbitrary positions. The limitation can be alleviated using the self-assembled plasmonic platform approach[19]; however, generation of heat and the resulting convective flow is unavoidable in plasmonic effect. This phenomenon may be utilized for concentration and patterning of sub-micron particles and DNAs[20], but compromises the capability of single particle manipulation.

In addition to increasing the force of an optical trap, significant research efforts have been made to increase the functionality of optical tweezers. One area that has been of particular interest is rotation and alignment of non-spherical particles. The alignment of particles is of special interest to applications such as tissue engineering, cell surgery, and nanofabrication. In conventional optical traps non-spherical particles tend to align their long axis to the direction of laser beam propagation, thus limiting the ability of aligning the particles in the plane of the substrate. Several methods have been used to circumvent this phenomenon. Using exquisite optical setups, a pair of closely separated optical traps can be generated by a spatial light modulator to hold different part of the particle[21], and a spiral interference pattern can be generated by combining a Laguerre-Gaussian beam with a plane wave with controllable optical path length difference to rotate non-spherical particles[22]. Utilizing optical torque birefringent particles can be rotated or spun by controlling polarization of the incident beam[23, 24]. However, in the primary applications of optical trapping, namely micro/nano-particle assembly and biological particles manipulation, there are limited number of samples with birefringence. The need for modifying and attaching biological samples to birefringent particles would involve perturbation to the biological specimens and increase sample preparation time, which is undesirable for biological studies that require high throughput. The rotation of regular dielectric particles can be achieved using polarization control of a Gaussian beam in optical tweezers[25], but with extremely high optical intensities.

In this paper, we propose enhanced optical manipulation utilizing the interaction of a polarized laser beam with 1D photonic crystal nanostructures. The approach uses a simple optical configuration as shown in Fig. 1a. We demonstrate trapping of sub-micron particles with low intensity while maintaining the full freedom of motion of traditional optical tweezers. Minimum trapping intensity for static trapping and trap efficiency for dynamic trapping (to trap and move particles simultaneously) were characterized for particles with a wide range of sizes. In addition, we show that rotation and alignment of non-spherical dielectric particles and cells can be achieved by controlling the polarization of the incident beam with low intensity. This approach can be expanded to a 2D photonic crystal structure to achieve fully controlled rotation. Although the current 1D photonic crystal structures are not used in the same sense as the broadly-known photonic bandgap configuration (the laser beam incidents normal to the grating in our configuration instead of along the periodic grating direction), we chose to use this terminology for its dominant use in describing periodic structures with periodicity around optical wavelengths [26]. Through FDTD simulations, we propose that the enhanced optical field near the nanostructure surface contributes to the enhanced trapping force, especially for sub-micron particles; and optical diffraction by the periodic structures results in orientation control, an effect more pronounced for micron-size non-spherical particles.

The effect of the photonic crystal structure on optical trapping force was investigated by first determining the intensity distribution after the incident light is scattered off the nanostructure. When coupled with appropriate material models, FDTD solvers can provide accurate representations of a wide range of optical behaviors. We calculated the intensity distribution in the volume surrounding the Gaussian beam spot incident on the surface of an aluminum grating with a period of 417 nm, same as that used in our experiments. A dispersive Lorentz-Drude model was used in conjunction with FDTD to accurately recreate the dielectric effects of aluminum, and it was verified that plasmonic effect was minimal. Once the intensity distribution is obtained, the potential energy of the optical trap for a particle versus location can be determined by convolving the volume of the particle through the intensity distribution, the gradient force can then be calculated[27] (see Methods section). Figure 1b shows the intensity distribution at the surface of the nanostructure, illustrating enhanced optical field when the incident light is polarized perpendicular to the grating lines. The trapping potential for different particle sizes versus location of the particle are shown in Fig. 1c and 1d, with the insets showing the corresponding trapping potential on a flat aluminum surface. From the potential energy the force can be found by taking the gradient of the trapping potential. Using this approach, the trapping force enhancement over a flat aluminum surface for 1 μm-size particles was found to be 1.85, while for smaller 350 nm particles the enhancement was found to be 10.7.

The experimental setup for characterizing the photonic crystal optical tweezers is described in the Methods section. Trapping forces were characterized in two ways. First trapped polystyrene beads of sizes with diameters from 190 nm to 10 μm with refractive index = 1.45 were dragged through solution (water, refractive index = 1.33) at varying speeds by moving the microscope stage relative to the laser spot. The laser intensity was then adjusted to find the minimum intensity at which the trap could hold the bead. From the velocity, trapping force was determined using Stoke's Drag Equation, $F_{drag} = -6\pi\eta r v_{flow}$, where $\eta$ is the viscosity of the surrounding medium, in this case water, $r$ is the particle radius, and $v_{flow}$ is the velocity of the fluid relative to the particle. Faxen's Law is taken into consideration to account for the particle proximity to the substrate. This measurement yielded trap efficiency in units of force per peak optical intensity. The result is shown as the square data points (blue curve) in Fig. 2a. On the average, the trap efficiency is about 20 times higher than what's reported using metal nanodots optical tweezers[13]. The asymmetry in the optical trap is demonstrated by the inset polar plot for trap efficiency. Trapping was also characterized by finding the minimum intensity at which the trap could overcome Brownian motion to hold a particle steadily. The result for various particle sizes is shown as the diamond data points (red curve) in Fig. 2a. For larger particles the Brownian motion was not noticeable. Figure 2b-d demonstrates trapping of a 590 nm-diameter fluorescent particle. The minimum intensity to maintain static trapping was found to be 34 $\mu W/\mu m^2$. Supplementary Movie 1 shows that the 190-nm particle was trapped initially; as the laser intensity is reduced, the particle moved away due to Brownian motion. Supplementary Movie 2 shows trapping and moving the 190-nm particle. The laser beam was turned off briefly initially to show the particle, turned on to trap and move the particle, then turned off again to show the particle at its final location.

As study of individual cancer cell nuclei may reveal informative data for cancer research[28], and holding the nuclei non-invasively with high reconfigurability is desirable to facilitating diagnostic applications, we performed trapping experiments for ovarian cancer cell nuclei using the photonic crystal optical tweezers. The nuclei were isolated and surface treated with bovine serum albumin to prevent clumping (see Methods section). Figure 2e-g show the snapshots of trapping a fluorescent ovarian cancer cell nucleus. The nuclei had a diameter of approximately 3 μm. The minimum intensity required to initiate trapping was characterized to be 16 $\mu W/\mu m^2$.

In addition to low intensity, two distinct trapping phenomena were observed. First for sub-micron particles the measured trapping efficiency (Fig. 2a) has a maximum at 750-nm particle diameter. Second, at larger particle sizes noticeable polarization dependence was observed. In the polarization state that produced the maximum diffracted field, perpendicular to the grating lines, the trap was found to be much stiffer when the particle was translated along the distribution of the diffracted modes. When the particle started outside the beam spot, it was repelled if approaching the trap across the grating lines, while attracted if approaching along the grating lines (see supplementary Movie 3). This suggests scattering force dominates for this phenomenon. The same effects were not observed with polarization parallel to the grating lines, where the diffracted field was much weaker. The observations imply that the trap asymmetry and corresponding enhancement was enabled by diffraction from the grating, which is a far-field effect and therefore more significant for larger particles.

Utilizing the polarization dependence, rotation and alignment was demonstrated for oblong 6.8 μm polystyrene beads (Fig. 3a-f and supplementary Movie 4) and ellipsoidal Listeria cell with a long axis of ~2 μm and a short axis of less than 1 μm (Fig. 4). In general, any non-spherical particle could be rotated into a minimum energy position as long as the intensity was sufficient to trap the particle. The 6.8 μm oblong beads could be rotated with intensity as low as 9.2 μW/μm$^2$, which to our knowledge is over four orders of magnitude lower than previously reported for polarization-defined rotation of dielectric, non-birefringent particles using optical tweezers[25]. The rotation speed versus optical intensity (Fig. 3g) was characterized with the long axis of the oblong beads initially close to 45º off the grating lines. Rotation speed near 35 degree/second can be achieved with sufficient intensity. The large uncertainty at high rotation speed was due to the variance in the beads' initial orientation, which affects the optical torque and initial speed of rotation significantly (see Methods). Figure 4 shows alignment characterization of the Listeria cells versus optical intensity using a 50x objective lens. When using a 20x objective lens, the minimum intensity for alignment could be reduced to 17 μW/μm$^2$, comparable to the local intensities reported for trapping and aligning *E. Coli* in patterned optical antennas (10-100 μW/μm$^2$) [16].

Although the current experiments were conducted using 633-nm wavelength, the principle of the photonic crystal optical tweezers can be readily applied to other wavelengths that are more bio-compatible by revising the design of the photonic crystal structures. The advantages of low power and high functionality of this approach are still the same, which is highly desirable for biological studies.

**Methods**

**Trapping potential calculation:** Once a three-dimensional electric field distribution is obtained, the spatially resolved trapping potential of that field can be determined by convolving the volume of a particle over the volume of the field as such[27]

$$U_{grad}(x,y,z) = \frac{n_2^2 - n_1^2}{16\pi} \int_v |\vec{E}(x,y,z)|^2 d\mathbf{v},$$

where $U_{grad}$ is the gradient trapping potential, $v$ is the volume of the particle, $n_1$ and $n_2$ are the indices of refraction of the surrounding medium and the particle, respectively. Once the potential is found the trapping force can be determined by calculating the gradient of the potential as follows,

$$\vec{F}_{grad}(x,y,z) = \nabla U_{grad}(x,y,z).$$

**Experimental setup:** A 633-nm HeNe laser with full output power of 35 mW was used as the light source for trapping and rotation. The laser beam was coupled into the optical path of a Zeiss Axioimager fluorescence microscope using a beam splitter. 50x, 20x, or 10x objective lens (N.A. = 0.55, 0.22, and 0.25, respectively) of the microscope was used for focusing the laser beam, producing observed spot diameters of approximately 3, 8 and 18 μm, respectively. The intensity of the laser beam was controlled using a neutral density attenuator, and the laser polarization was controlled using a rotating half-wave plate. The intensity of the optical beam reported in this paper was obtained by measuring the optical power after the microscope objective lens, divided by the laser beam spot size which can be estimated from the CCD camera images. Samples were generally observed using the CCD camera with the microscope operated in bright-field mode. For very small particles and biological samples, fluorescence imaging was used. In bright-field mode a regular beam-splitter not optimized for the laser wavelength was used and a maximum optical power of 9 mW could be delivered to the sample; whereas in fluorescence mode a dichroic beam-splitter was used and 27 mW maximum power could be delivered.

The substrate used for the current experiment was a holographic (sinusoidal profile) diffraction grating with a period of 417 nm and groove depth of 150 nm. The grating surface was aluminum. Test particles were suspended in water maintained above the grating surface using a glass coverslip and spacers.

**Nuclei preparation:** Nuclei were isolated from the ovarian cancer cell line 2008. The cells were washed twice in ice cold phosphate buffer solution (PBS), re-suspended in a previously described hypotonic lysis buffer[29], and homogenized in a dounce homogenizer for one minute. The cellular lysate was layered over a 1.8 M sucrose solution and ultra-centrifuged for 2.5 hours at 23,600g to remove cytoplasmic material. The resulting nuclear pellet was resuspended in 1% bovine serum albumin (BSA) in PBS and stained with the DNA dye, acridine orange which has an emission maximum at wavelength = 525 nm.

**Rotation speed characterization:** The rotation speed characterization began with targeting a 6.8um oblong polystyrene bead that is close to 45º out of alignment with the grating. If no bead can be found that is 45º out of alignment, a bead that is closer to parallel with the grating was targeted. The bead was trapped by approaching it with the laser spot quickly. If the bead was approached slowly, it would be pulled into the laser spot from a large distance away, giving it an initial angular momentum which might make the bead appear to align more quickly than it would otherwise. The trapping and rotation process was video-recorded and the rotation speed was determined by measuring the time duration for the bead to rotate from 45º to a steady position perpendicular to the grating in the video. If the initial position of the bead was greater than 45º, the time duration was determined by starting to count video frames when the bead reached the 45º position. The initial position was determined by the angle the oblong bead makes with the grating when it was first trapped by the laser, not the position prior to trapping. The measurement was repeated five times for each laser intensity tested, and the average and standard deviation was determined.

**Acknowledgements**

This work has been supported by the National Science Foundation (DBI 0454324) and the National Institute of Health (R21 EB005183). T. Mentele and S. Bachar thank the Amgen Scholar Program for the support. E. Knouf was supported in part by Public Health Service, National Research Service Award, T32 GM07270, from the National Institute of General Medical Sciences. M. Tewari acknowledges New Development funding from the Fred Hutchinson Cancer Research Center, a grant from the Mary Kay Ash Foundation and a Career

Development Award from the Pacific Ovarian Cancer Research Consortium (POCRC) Ovarian SPORE (NIH grant P50 CA83636).

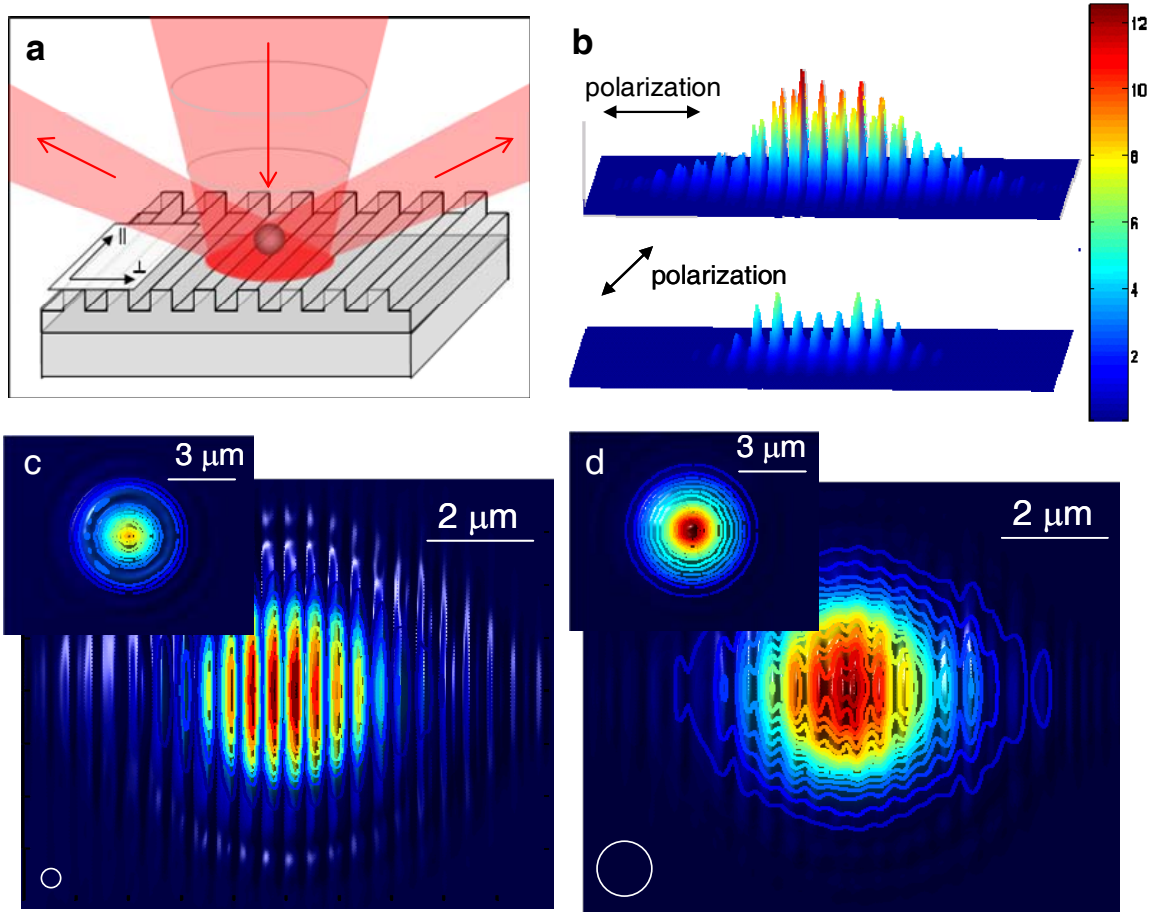

**Figure 1. Enhanced optical trapping utilizing 1D photonic crystal nanostructures. a,** Schematic drawing of the approach. The incident beam is diffracted by the periodic nanostructure at far field. **b,** The intensity distribution of light with two orthogonal polarizations at the surface of an aluminum grating with a 417 nm period obtained using FDTD simulations. The distribution is normalized to the intensity on a flat aluminum surface. **c** and **d,** Trapping potential for particles directly above the grating surface versus location of the particle for **c,** a 350 nm polystyrene bead and **d,** a 1 μm polystyrene bead. The white circles illustrate the sizes of the particles. Insets show the trapping potential above a flat aluminum surface for the same particle size as comparisons. The values are normalized for each particle size. For all FDTD simulation figures the field of view is $10 \times 8$ μm$^2$.

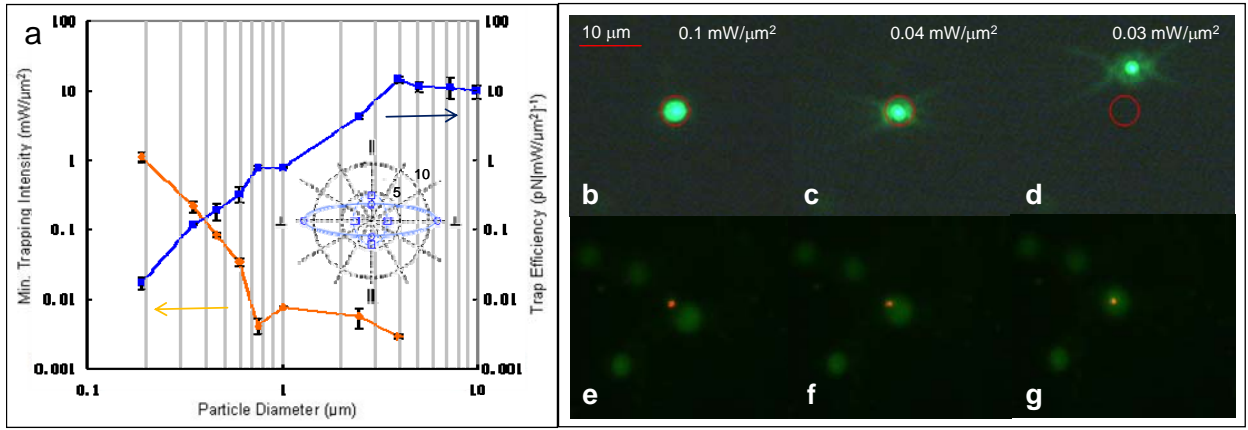

**Figure 2. Trapping characterization of the photonic crystal optical tweezers. a,** Trap efficiency and minimum trapping intensity measured for polystyrene beads of various sizes with beam polarization perpendicular to grating lines. Inset shows trap asymmetry in trapping efficiency for translating a 3.87 um polystyrene bead perpendicular and parallel to the rules of the grating. The solid line (large asymmetry) is obtained with incident light polarized perpendicular to the grating, and the dash line (small asymmetry) is obtained with incident light polarized parallel to the grating. The unit is in $(pN[mW/\mu m^2]^{-1})$. **b-d,** Trapping demonstration of a fluorescent 590 nm polystyrene bead. The red circle indicates the position of the laser spot as the laser light was too dim to be seen. At first the particle is trapped within the spot at higher power, as the power is lowered the Brownian motion of the particle overcomes the trapping force, allowing the particle to escape. **e-g,** Trapping demonstration of a fluorescent ovarian cancer cell nucleus. The minimum intensity required to initiate trapping was 16 $\mu W/\mu m^2$ obtained using a 20x objective lens.

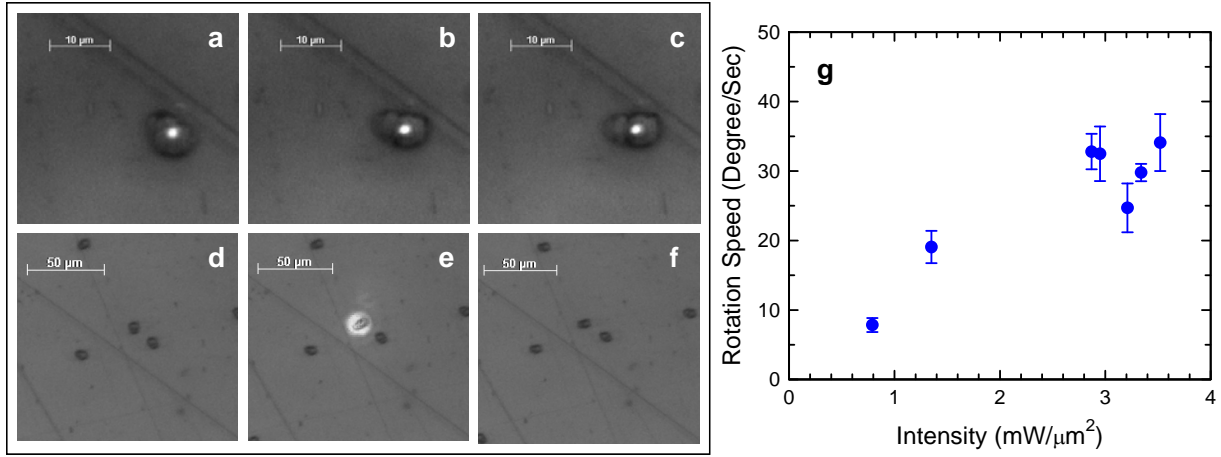

**Figure 3. Rotation and alignment of non-spherical particles using photonic crystal optical tweezers. a-c,** Rotation of an oblong polystyrene particle on an aluminum grating under the illumination of a laser beam polarized perpendicular to the grating lines. The laser beam was focused with a 50x objective lens. **d-f,** The oblong polystyrene particle was aligned with its long axis perpendicular to the grating line after being illuminated by the polarized laser beam. The focusing of the laser beam and image-taking was through a 10x objective lens. **g,** Characterization of rotation speed versus optical intensity. Rotation speed approaching 35 degree/sec can be achieved with sufficient laser intensity.

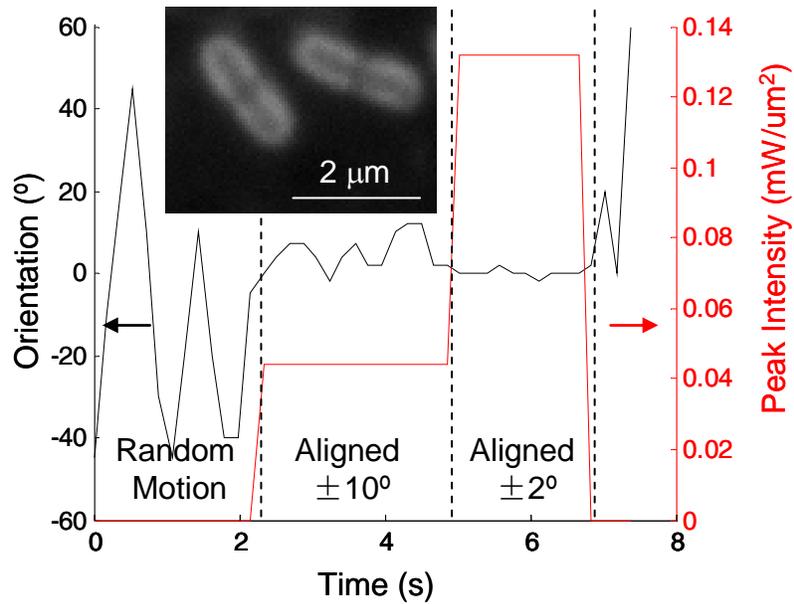

**Figure 4. Characterization of Listeria cell (inset) alignment using photonic crystal optical tweezers.** The orientation angle of the Listeria cell was recorded as the laser intensity was adjusted. As laser power increases the cells are more stiffly aligned perpendicular to the grating lines. This measurement was performed using a 50x objective lens. The cell could be aligned to within ±10º when the laser intensity exceeds 40 μW/μm$^2$.